**Improving the Environmental Perception of Autonomous Vehicles using Deep Learning-based Audio Classification**

**Finley Walden**
Undergraduate Student
Department of Computer Science
The University of Alabama
3041 Cyber Hall, Box 870205, 248 Kirkbride Lane, Tuscaloosa, AL 35487
Tel: (850) 748-4621; Email: fswalden@crimson.ua.edu

**Sagar Dasgupta\***
Ph.D. Student
Department of Civil, Construction & Environmental Engineering
The University of Alabama
3014 Cyber Hall, Box 870205, 248 Kirkbride Lane, Tuscaloosa, AL 35487
Tel: (864) 624-6210; Email: sdasgupta@crimson.ua.edu

**Mizanur Rahman, Ph.D.**
Assistant Professor
Department of Civil, Construction & Environmental Engineering
The University of Alabama
3015 Cyber Hall, Box 870205, 248 Kirkbride Lane, Tuscaloosa, AL 35487
Tel: (205) 348-1717; Email: mizan.rahman@ua.edu

**Mhafuzul Islam, Ph.D.**
Senior Research Scientist
General Motors R&D,
Michigan, USA
Tel: (864) 986-5446; Email: mdmhafuzul.islam@gm.com

\*Corresponding author

## ABSTRACT

Sense of hearing is crucial for autonomous vehicles (AVs) to better perceive its surrounding environment. Although visual sensors of an AV, such as camera, lidar, and radar,  help to see its surrounding environment, an AV cannot see beyond those sensors' line-of-sight. On the other hand, an AV's sense of hearing cannot be obstructed by line-of-sight. For example, an AV can identify an emergency vehicle's siren through audio classification even though the emergency vehicle is not within the line-of-sight of the AV. Thus, auditory perception is complementary to the camera, lidar, and radar-based perception systems. This paper presents a deep learning based robust audio classification framework, aiming to achieve improved environmental perception for AVs. The presented framework leverages a deep Convolution Neural Network (CNN) to classify different audio classes. UrbanSound8k, an urban environment dataset, is used to train and test the developed framework. Seven audio classes—i.e., air conditioner, car horn, children playing, dog bark, engine idling, gunshot, and siren, are identified from the UrbanSound8k dataset because of their relevancy related to AVs. Our framework can classify different audio classes with 97.82% accuracy. Moreover, the audio classification accuracies with all ten classes are presented, which proves that our framework performed better in the case of AV-related sounds compared to the existing audio classification frameworks.

# INTRODUCTION

Autonomous vehicles (AVs) will roam around your neighborhood in the near future, and they will replace conventional human-driven vehicles. It is expected that AVs have the ability to increase safety significantly compared to human-driven vehicles as AVs could eliminate human errors. In order to ensure safe AV operation, the vehicle must have a robust environmental perception and understanding of its surroundings. The perception system of AVs must be able to detect, identify and track surrounding both static and moving objects in real-time. Currently, AV's perception system is based on multimodal sensing, and the sensor array consists of different types of cameras, long and short-range radars (Radio Detection And Ranging), LiDAR (Light Detection and Ranging), ultrasonic transducers, and GPS (Global Positioning System) receiver. AV uses data from all these sensors and creates a perception model of the surrounding environment. Nevertheless, the shortcoming of these sensors can make the AV vulnerable. For example, limited line of sight of these sensors and limited functionality during extreme weather conditions can lead to faulty perception and lead to a crash.

Researchers are aiming to replace the human driver with an AV's built-in advanced decision-making system that mimics a human brain to overcome human driving errors. For a human driver, surrounding sounds play a vital role in driving-related decision-making, such as reacting to honking or fast-approaching car sounds. However, all the AV manufacturing companies except Waymo exclude audio sensors from the environmental perception module. AVs also need to be aware of the surrounding sound's make better decision making. For example, an AV can identify an emergency vehicle's siren through audio classification even though the emergency vehicle is not within the line of sight of the AV. Hence, detection and classification of the audio of the surroundings are a necessity for a robust level 4 and level 5 autonomy because a vehicle that cannot hear its surroundings will not be able to react to horns, sirens, breaks screeching, and other common sounds that human driver's sense and utilize their driving decision making.

The audio classification acts as a complementary module for AV's perception system. For urban environment modeling and interpretation, deep learning algorithms can be trained with annotated urban environment-related sounds to classify different types of audio, such as sirens from emergency vehicles and children playing in a residential area. Such classification can be further fed to AV's advanced decision-making system to make a better driving decision. If an AV is roaming around a residential area or traffic calming zone and due to visual obstructions, such as walls and trees, the AV's field of view is minimized. As a result, there is a possibility that children are playing around and may suddenly come in front of the AV, and if there isn't enough time to react, an accident can happen. On the other hand, suppose an AV can detect sounds related to children's playing. In that case, an AV can make a robust decision while driving, maintaining lower speed and greater caution to get more time to respond to the sudden change in the surrounding environment.

In this paper, a robust audio classification framework is presented in the context of AV. Specifically, the primary contribution of this paper is to develop a convolution neural network (CNN) architecture to classify different audio types with higher accuracy compared to the classification accuracy from existing literature. The performance of the CNN architecture is evaluated using the UrbanSound8k dataset (*1*), which includes 8732 annotated urban sounds of 10 different classes. As CNN takes an image as an input, the processed audio files are converted to images using Mel Frequency Cepstral Coefficients (MFCCs). Moreover, in order to compare the performance of the developed CNN with existing models, the CNN model is further trained and tested with all ten classes of UrbanSound8k.

## RELATED WORKS

The process of sound classification through machine learning is a developed topic that has been applied for many different practical purposes. There are various models and experiments with tangible results, and many of these models can potentially augment environmental awareness in AVs. While there are many different machine learning techniques to classify sounds, most experiments opt for deep learning approaches over classical machine learning strategies. In (*2*), the authors compare the performance of classical machine learning models with the performance of deep learning models , i.e., Convolution Neural Network (CNN) and Support Vector Machine (SVM). These comparisons are performed in the context of sound classification. Generally, deep learning models perform better. They are more efficient and are the widely used strategy for sound classification. The authors' classical model did perform well on ESC-50 (*3*) and UrbanSound (*4*) datasets; however, the model was not computationally efficient. Overall, the authors in (*2*) conclude that in terms of processing time, feature extraction is the most prevalent factor by far, so both classical and deep learning models perform similarly in the long run.

The model presented in (*5*) uses residual blocks in a CNN to achieve high accuracies on different datasets in the context of environmental sound classification. The authors of (*6*) discovered that a robust Vision Transformer (ViT) model for audio event classification performed well on a large dataset of YouTube sound clips. Overall, for general sound classification, many researchers and industry professionals extract spectrogram data from raw wave data in order to use deep learning techniques common in image classification to classify the sounds.

CNN is widely used for environmental and urban scene (image) classification. In (*7*), the authors used a model to classify audio scenes. Their Convolutional Recurrent Neural Network (CRNN) model which is a combination of CNN and Recurrent Neural Network (RNN) outperformed a standard CNN by around 17% in terms of accuracy rate, achieving an accuracy rate of around 91% for general audio scene classification. On the other hand, the authors of (*8*) used a resource adaptive CNN (RACNN) that achieved an accuracy rate of around 97% on the UrbanSound8k dataset. The accuracy varied with changing hyperparameters. This lightweight model still manages to attain a high accuracy despite having far less trainable parameters.

Other deep learning applications exist in sound classification, including using neural networks for feature extraction. In (*9*), the authors propose an alternative feature extraction method to popular existing models. Their deep CNN extracts features exponentially quicker than existing models and performed well in different contexts. Further CNN models for sound classification can be found, including a two-stream CNN based model for environmental sound classification (*10*). This model also obtained an accuracy rate of around 97% using the UrbanSound8k dataset, although with a far larger quantity of trainable parameters.

In the context of robotics, many researchers use audio to augment their machines' awareness of its surroundings. In (*11*), the authors attempt to use engine and motor noises to estimate the motion of a robot. They used a neural network to implement this, and their model improved their estimation of the motion by 26%. In (*12*) the authors explore the use of audio input to classify types of terrain. They argue that it is important for robotic equipment to be able to recognize its environment. Using a deep CNN, they achieved accuracy rates of above 90% for nine types of terrain.

The authors of (*13*) discuss a wide variety of applications of auditory perception in AVs. Their first focus outlines how vehicles might be able to classify sirens. They also explore applications, such as road-tire noise perception, vehicle classification, and several off-road applications of the use of audio input in AVs. The authors of (*14*) seek a resilient environmental scene and location classification through the use of sound. Due to the volatility of vision-based solutions due to impediments, such as weather and daylight, they think that using sound and other sensory input can augment the ability of a robotic vehicle to recognize its location and proceed correctly. They use a CNN to classify audio samples of different terrains, all at an accuracy rate of 98%.

Most of the research related to in-vehicle audio classification pertains to detecting emergency vehicles. In (*15*), the topic of emergency vehicle detection is explored. They tested a system with emergency sirens from a variety of different positions. While their success rate was not universally high, it was high when the sirens were within 50 meters, a critically close range. In (*16*), the authors work to classify sounds in the context of autonomous vehicles, particularly classifying vehicles as emergency vehicles. They attained a test accuracy of 88% percent using a combination of video and audio to classify the vehicles. The authors of (*17*) attempt to improve the classification of sirens in scenarios with background noise, and they attempt to improve the feature extraction through the use of k-means. Using a CNN, they classified sirens with an average accuracy rate of 94%. In (*18*), the authors use a two-stream CNN model to classify the horns of cars and trains. Their model outperformed several other models with an accuracy of 95.11% on a train horn dataset.

Compared to existing literature, this paper aims to develop a sound classification framework with a new CNN architecture to improve the classification accuracy of audio sound related to AVs. As per our knowledge, audio classification related to AVs is focused on only sirens of emergence vehicles. However, this paper evaluate the performance of audio classification accuracy for seven different audio classes related to AVs.

## DATASET DESCRIPTION

The UrbanSound8k dataset (*1*) consists of of 8,732 samples from urban environments, categorized into ten classes: air conditioner, car horn, children playing, dog bark, drilling, engine idling, gunshot, jackhammer, siren, and street music (see **Figure 1**). It includes audio samples and metadata. Each audio sample is a ".wav" file. The length of each audio file varies from others; however, the length of all audio files are less than five seconds. **Figure 2** represents sample raw wav files of five classes, i.e., air conditioner, car horn, children playing, dog bark, and engine idling. Out of all the samples presented, only the dog bark sample is mono—i.e., it has just one channel. The remaining presented samples are stereo—i.e., the audio files have two channels. The metadata includes the audio file name, source ([www.freesound.org](www.freesound.org)) (*19*), audio file ID, class ID, occurrence ID, and slice ID that can be used to identify the occurrence from which the audio file is sliced. All ten classes give information on the sensor's surroundings from which the sound data is extracted. Among all ten classes, drilling, jackhammer, and street music classes are not particularly important for AV's better environmental perception. Hence, these three classes are excluded from the model training and testing for the AV-specific application.

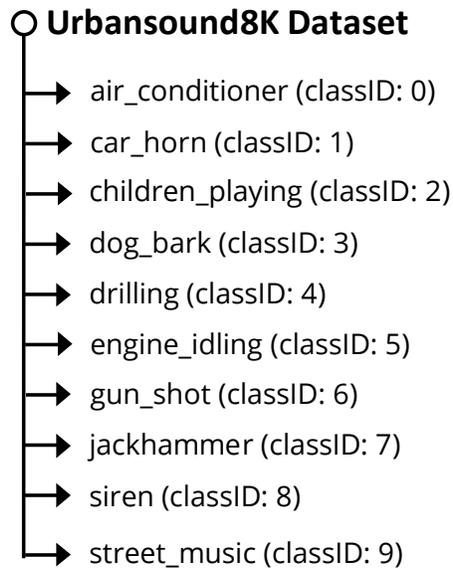

**Figure 1 Urbansound8k dataset classes**

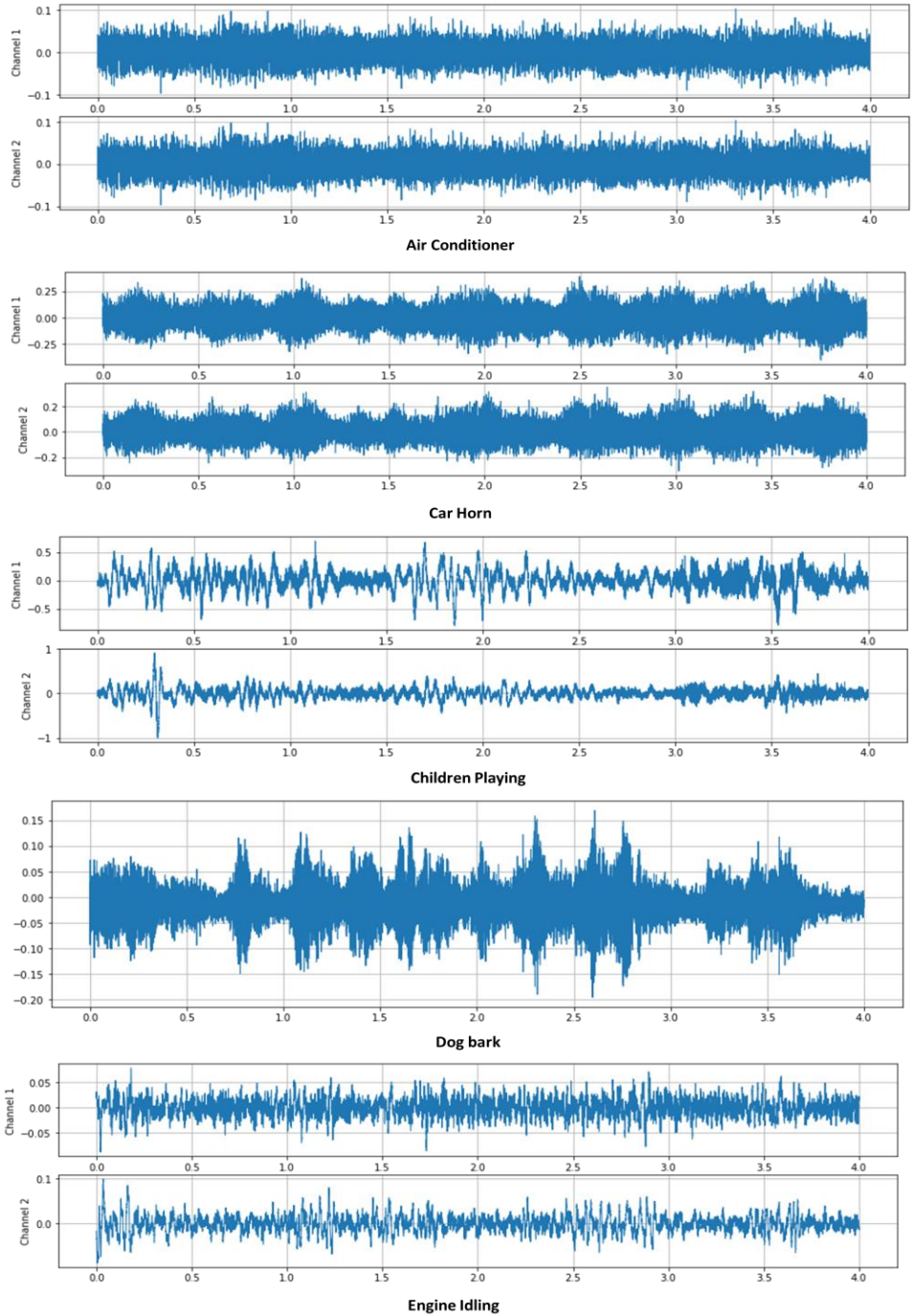

**Figure 2 Urbansound8k raw sample waveforms**

## METHOD

In this section, we presented the audio classification pipeline used in this study and compared several machine learning models in classifying the urban audio data. From existing literature, we found that convolution neural network (CNN) and long-short term memory (LSTM) are widely used for audio data classification. Therefore, we have used our audio classification pipeline to compare between our developed ML model and existing ML models. The audio classification pipeline, as shown in **Figure 3**, contains four basic blocks, i.e., dataset, splitting the dataset into training and validation, data preprocessing, and CNN audio classification framework.

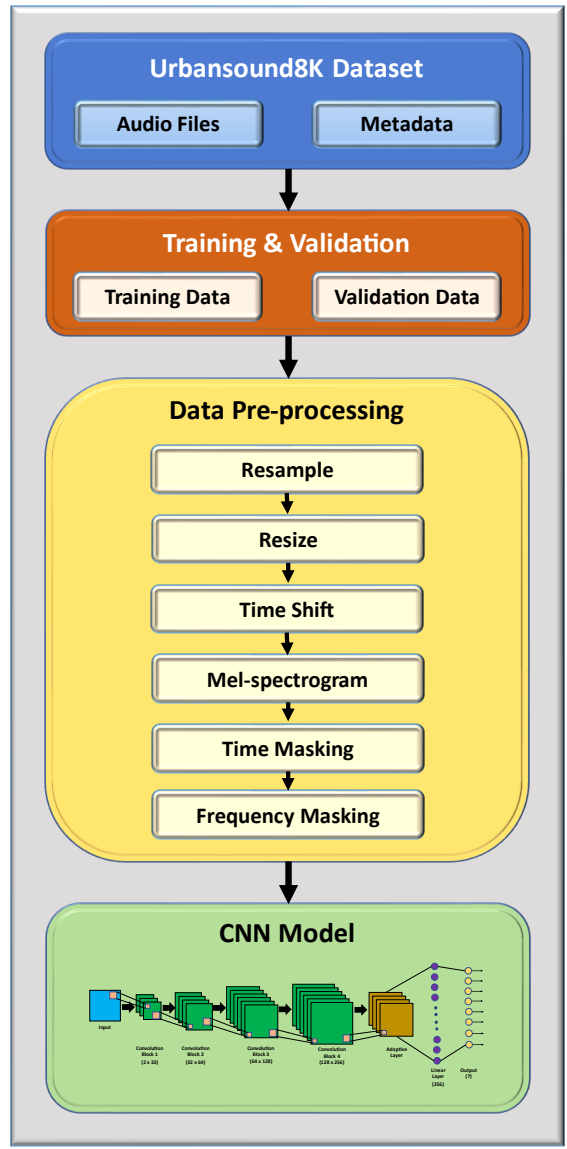

**Figure 3 CNN-based Audio Classification Pipeline**

## Data Preprocessing

The purpose of the data prepressing step is to create all the data into a unified format. In the Urbansound8k dataset, each sound is in a waveform format. In the first step, the data pre-processing block reads these waveforms formatted files. To ensure that all the samples are in stereo format, the audio is then resampled to 44,100 Hz and rechaneled to two channels. The resampled waveform is then padded or truncated to a specific fixed length. This process standardizes the data's length, sampling rate, and number of channels, ensuring that each feature has the same dimensions when extracted. The final step in the data preprocessing and standardization process is a time shift by a factor of 0.4, which is a data augmentation method equivalent to rotating or scaling images. This step ensures that the model can better generalize and make predictions for a larger variety of data for each given class.

**Feature Extraction**

Operating on the preprocessed waveform data, the program converts the audio into a Mel spectrogram, enabling it to be processed by an image classification neural network. Since the samples have the same dimensions, the Mel spectrograms are all the same size, eliminating the need for further standardization. After generating the Mel spectrogram from the waveform, we perform data augmentation on the spectrograms. First, certain frequency ranges are masked. On a Mel spectrogram, this equates to horizontal bars over the plot, masking the values of randomly generated frequencies. A similar process occurs with time masking, which appears as vertical bars across specific time periods. Both masking steps mask up to 10% of their given dimension, ensuring that the data is still readable and that a machine learning model is capable of finding patterns consistently. The purpose of this data augmentation is to prevent overfitting and ensure that the model finds general patterns in the spectrograms, as opposed to overestimating the importance of specific training data features that will prevent generalization and cause overfitting. After data preprocessing, feature extraction, and data augmentation, each audio sample has two spectrograms of size 64x344. The two spectrograms result from the rechanneling to two-channel stereo specifications during the preprocessing step. **Figure 4** shows a sample visual representation of raw 4 seconds of audio data **(Figure 4(a))** and its corresponding spectrograms **(Figure 4(b)).** Finally, the data is normalized by **Equation 1**:

$$X = \frac{X - \mu}{\sigma} \qquad (1)$$

where, $X$ is the input amplitude, $\mu$ is the mean of the input, and $\sigma$ is the standard deviation of the input. This enables the model to ignore variable differences and classify samples solely based on a normalized representation of the spectrogram data.

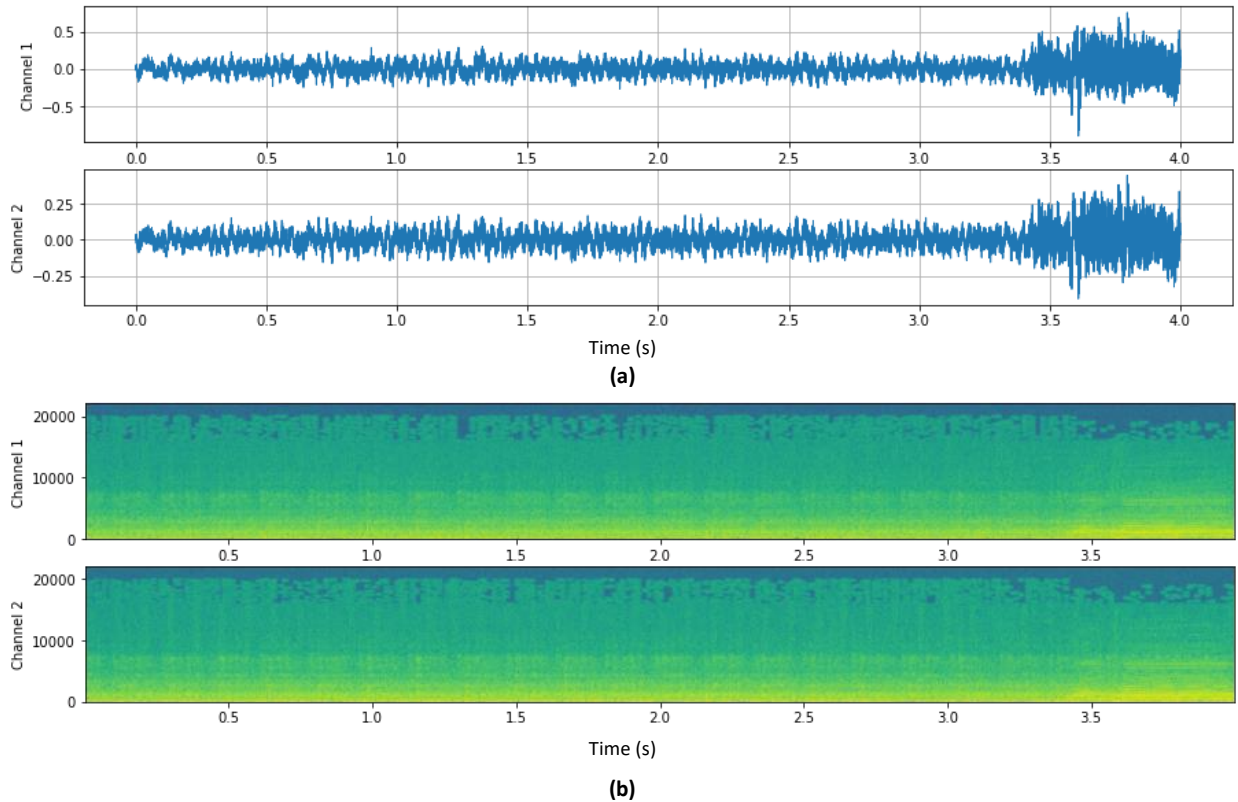

**Figure 4 Visual display of audio data before (a) and after (b) converting to spectrogram**

**Audio Classification Model: Convolutional Neural Network (CNN)**

After creating the normalized Mel spectrogram images, we use a convolution neural network (CNN) to classify the audio data. **Figure 5** presents a CNN architecture that utilizes four convolution blocks. The hyperparameters includes: Rectified Linear Unit (ReLU) as activation function, batch normalization, the Adam optimizer, and 2-D Average Pooling. This CNN model extracts the features through the convolution blocks, and classify the data into one of the 10 classes using fully connected (FC) layers. The model as a whole contains approximately 2.1 million trainable parameters. **Table 1** provides the details of all the hyperparameters used in the construction of the convolution blocks. The convolution layers' hyperparameters are arranged to maintain a reasonable number of total parameters while enabling a powerful classification model. When tested with feature extraction filter sizes of 8, 16, 32, and 64, which is one of the primary hyperparameter of the convolution layer, our accuracy rate suffered by 8%. Therefore, this larger model was used that decreases the training and classification efficiency without dramatically increasing the accuracy of the model. The padding and stride added further enhance the model's classification abilities by enabling the model to have a larger number of layers. Padding refers to addition of pixels (adding layers of zeros) to input images to preserve the border information of the image after using filter during a convolution operation. Whereas stride is a CNN filter parameter which detects the amount of movement over the input image. The ReLU activation function was the suitable choice for our model, which improved the classification accuracy further.

Finally, we normalized the audio inputs and include a batch normalization function after the activation function on each convolution block to reduce the training time.

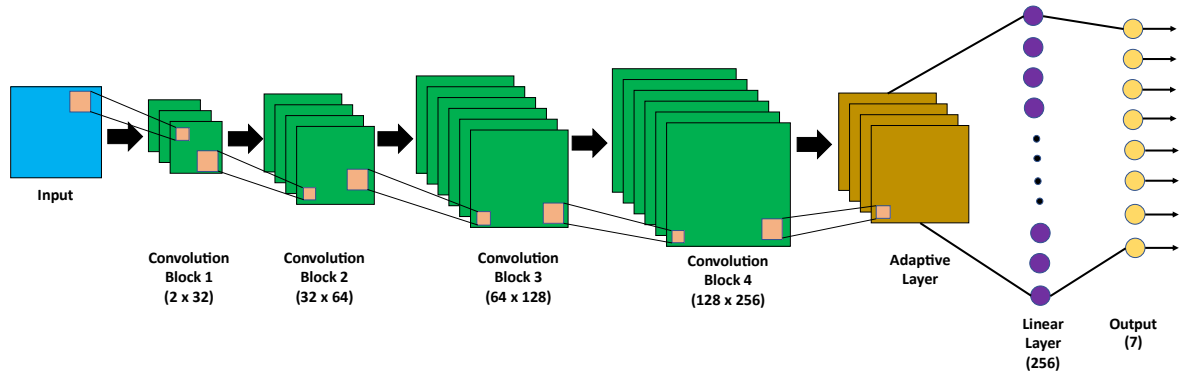

**Figure 5 CNN Architecture used for audio classification**

**TABLE 1 Summary of CNN Layers**

| Convolution Block | Filter size | Kernel Size | Padding | Stride |
|---|---|---|---|---|
| 1 | 32 | (3, 5) | (2,2) | (2,2) |
| 2 | 64 | (3, 5) | (2,2) | (1,1) |
| 3 | 128 | (5, 5) | (2,2) | (1,1) |
| 4 | 256 | (5, 5) | (2,2) | (1,1) |

## EEVALUATION RESULTS AND DISCUSSIONS

In this section, we perform two analyses based on our developed pipeline and CNN model. In the "Analysis 1," we took consider a use case for autonomous vehicle (AV), and in Analysis 2, we evaluate the existing models with our developed CNN model to compare the accuracy of the audio classification.

### Analysis 1: Autonomous Vehicle Application

Due to our goal of classifying sound in the context of an autonomous vehicle, we opted to train the model with only seven of the ten classes in the UrbanSound8k dataset: air conditioner, car horn, children playing, dog bark, engine idling, gunshot, and siren. After an 80/20 train/test split and 100 epochs of training, the model produced a training accuracy of 100% and a testing accuracy of 97.82%. **Figure 6** depicts the loss function and accuracy value over time. The x-axis is the number of epochs, the primary y-axis is the loss, and the secondary y-axis is the accuracy of the prediction. The loss in pytorch is calculated using cross entropy loss. Increased training epochs reduce the loss and stabilize at around 100 epochs, while training accuracy improves.

According to the confusion matrix (**Figure 7**), air conditioning has a 98.94% accuracy, air conditioning sounds are misclassified as car horn and dog bark for 0.53% and 0.53% times, which are insignificant. Car horn has an accuracy of 97.12% and misclassified with dog bark and gunshot classes; however, the false positivity is not high. Children playing also have high accuracy rate of 98.40% which is slightly less than air conditioning. The dog bark class is clearly confused at a high rate; however, the consequences of this confusion are largely insignificant in most cases. The siren class has the highest rate of accuracy 98.97% with minor rate of confusion with dog bark and gunshot. Engine idling sound detection has an accuracy of 98.11% and it is falsely labeled as air conditioning, dog bark and gunshot. The last class gunshot has a classification accuracy of 98.59% and it is falsely predicted as dog bark at a nontrivial rate. Overall, the model is capable of classifying 7 audio classes with more than 98% accuracy for all classes except dog bark class.

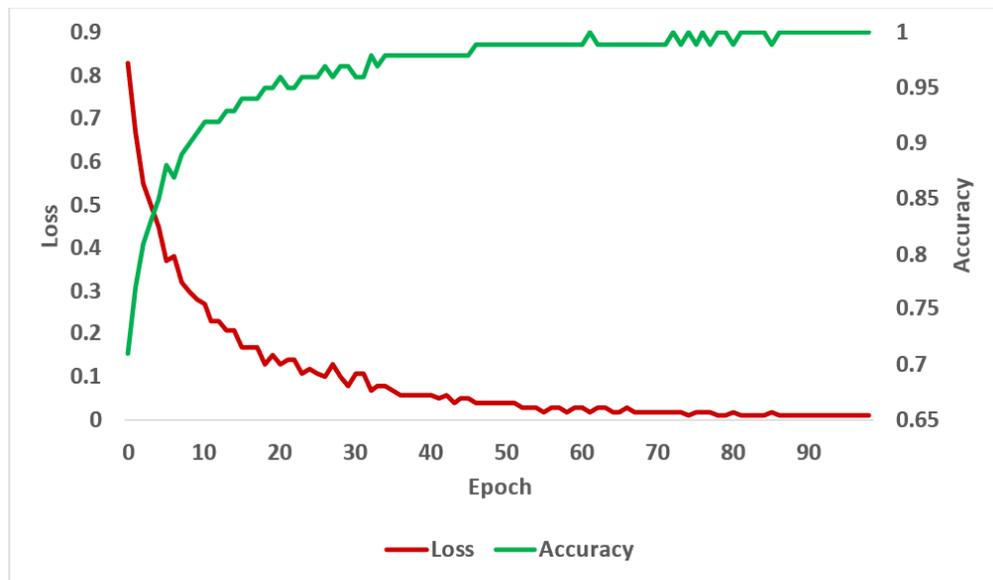

**Figure 6 CNN Training Performance (Analysis 1)**

The model's strengths lie in its overall high accuracy. In the context of autonomous vehicles, this model would be useful for the recognition of a vehicle's surrounding environment. While all the classes included in the UrbanSound8k Dataset are useful, the fact that the model performs so well on car horn, children playing, and engine idling is a major strength of the model in the context of autonomous driving. These three classes carry heightened importance: recognizing car horns is vital in a context with robotic and human drivers; the recognition of children playing and general human noises could prevent fatal accidents; and the ability to recognize an idling engine can enhance an AV's ability to recognize stopped vehicles in the vicinity.

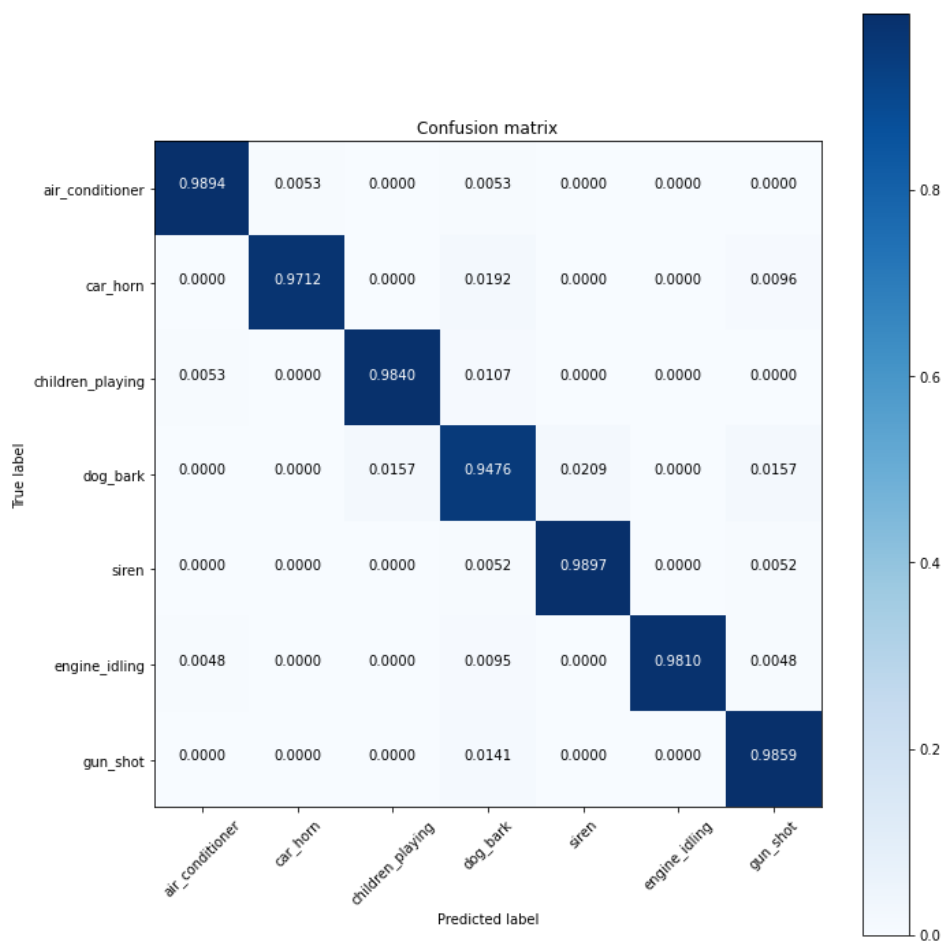

**Figure 7 Confusion matrix of CNN in the context of an autonomous vehicle application (Analysis 1)**

**Analysis 2: Comparison with Existing Models**

In this analysis, we consider all 10 classes that can be classified using ML models. First, we split the dataset into an 80/20 train-test split. After 100 epochs of training the model produced a training accuracy of 99% and a testing accuracy of 96.4%. The confusion matrix for this analysis is shown in **Figure 8.** Three out of the ten classes—i.e., air conditioner, car horn, and gunshot—attained a result of 99%, with children playing and engine idling having an accuracy of 98%. The two weakest classes were drilling at 94% and street music at 93%. As the confusion matrix demonstrates, air conditioning and car horn both have high accuracy rates of 99%. Each class is incorrectly selected over another class at a rate of 1%. Children playing has four instances of nontrivial rates of false positivity, with street music being incorrectly identified as children playing at a rate of 3%. The dog bark has similar results to children playing, although its false negative rate is slightly higher. The drilling class and the jackhammer class are clearly confused at a high rate; however, the consequences of this confusion are largely insignificant in most cases. The engine idling class has high accuracy (98%), with a minor rate (1%) of confusion with drilling. The gunshot class is correctly identified with its accuracy of 99%, but it is falsely labeled as children playing with 1%. The siren class is falsely labeled as four other classes at a nontrivial rate, and street music was mistaken for a siren at a rate of 2%. Street music was falsely predicted as seven of the nine other classes, albeit at low rates for each one. Overall, the shortcomings of the model's predictive capabilities are most notably present in confusion between the drilling and jackhammer class, the false negative rate of the siren class despite its distinct sound, and the general low accuracy of the street music class.

As we compare the accuracy with the other existing ML models, our CNN model outperforms the presented model in the (*8*)and the (*10*) papers in a few key classes, most notably children playing and car horn. However, as shown in Analysis 1, this advantage of our model would be crucial in the context of autonomous vehicles due to the aforementioned situations. **Table 2** shows the performance of our proposed model compared with that of the Two Stream CNN (TSCNN) (*10*) and that of the RACNN from (*8*). The overall average does trail slightly in some of the sounds, such as engine idling, gun shoot siren, and street music, but for the other types of sounds our model outperforms in classifying/detecting the appropriate sound.

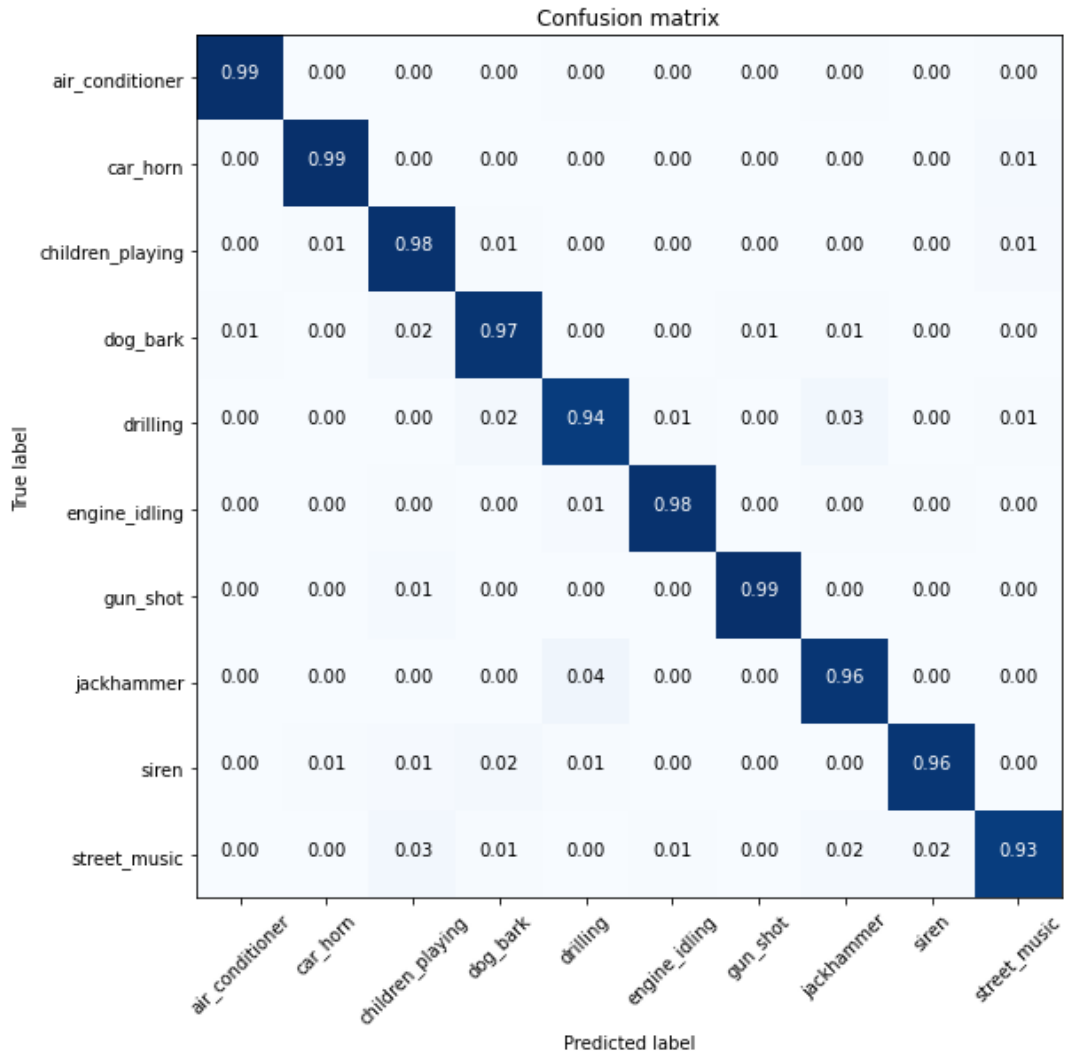

**Figure 8 Confusion matrix of Comparison with Existing Models (Analysis 2)**

**TABLE 2 Comparison with other models on individual classes in the UrbanSound8k dataset**

| Model | Air Conditioner | Car Horn | Children Playing | Dog Bark | Drilling | Engine Idling | Gun Shoot | Jack Hammer | Siren | Street Music | Average |
|---|---|---|---|---|---|---|---|---|---|---|---|
| TSCNN (*10*) | .99 | .94 | .97 | .95 | .97 | .99 | .95 | .97 | .99 | .97 | .972 |
| RACNN (*8*) | .98 | .98 | .95 | .96 | .98 | 1 | 1 | .98 | 1 | .95 | .975 |
| Our Model | .99 | .99 | .98 | .97 | .94 | .98 | .99 | .96 | .96 | .93 | .964 |

## CONCLUSIONS

In scenarios with poor weather, obstructed visual sensors, and urban or environmental obstructions, such as trees, construction, and buildings, access to other key sensory data will be crucial in ensuring that autonomous vehicle operation is safer than conventional, human-operated vehicle. Our CNN model would, in conjunction with sensory hardware attached to the vehicle, be able to enhance the environmental awareness of an AV. The evaluation results presented in this study can be useful in a variety of scenarios that could save lives or increase operational efficiency. For example, when driving through a residential area, an AV should have the ability to recognize audio from human sounds, as with the children playing class above. In a storm, a car approaching an intersection should have the ability to recognize a horn, an idle engine, and other key sounds that have the potential to dramatically reduce the likelihood of an accident. The accuracy rates of this model suggest that the model has a high enough confidence to affect the decision-making process of an AV. Our CNN's strengths, particularly with regard to classes useful to autonomous driving, make it a suitable candidate for future deep learning solutions to audio classification in this context.

## AUTHOR CONTRIBUTIONS

The authors confirm their contribution to the paper as follows: study conception and design: F. Walden, S. Dasgupta, M. Rahman, M. Islam; data collection: F. Walden, S. Dasgupta, M. Rahman; interpretation of results: F. Walden, S. Dasgupta, M. Rahman; draft manuscript preparation: F. Walden, S. Dasgupta, M. Rahman. All authors reviewed the results and approved the final version of the manuscript.